\documentclass[aps,prl,reprint]{revtex4-1}
\usepackage{graphicx}

\begin{document}

\title{Critical p=1/2 in percolation on semi-infinite strips}

\author{Zbigniew Koza}\email{zbigniew.koza@uwr.edu.pl}
\affiliation{Faculty of Physics and Astronomy, University of Wroc{\l}aw,
	50-204 Wroc{\l}aw, Poland}

\date{\today}

\begin{abstract}
We study site percolation on lattices confined to a semi-infinite strip.
For triangular and square lattices
we find that the probability that a cluster touches
the three sides of such a system at the percolation threshold has the continuous limit 1/2 and argue that
this limit is universal for planar systems.
This value is also expected to hold for finite systems for any self-matching lattice.
We attribute this result to the asymptotic symmetry of the separation lines
between alternating spanning clusters of occupied and unoccupied sites formed
on the original and matching lattice, respectively.

\end{abstract}


\maketitle

Introduced as a model of transport through a random medium \cite{Broadbent1957},
percolation has attracted attention as one of the simplest, purely geometrical
models with a phase transition.
In its basic version it is defined on a lattice with either nodes or edges
being chosen with some probability $p$ to be ``open'' to transport.
If one considers the limit of the system size going to infinity
(also known as the continuous or thermodynamic limit),
then  below some critical value $p_\mathrm{c}$ the probability
that the system as a whole is permeable is 0, whereas for $p > p_\mathrm{c}$
it is 1.

Several rigorous results have been obtained for percolation so far.
The concepts of matching and dual lattices were applied to predict the exact
values of $p_\mathrm{c}$ for site percolation on the triangular lattice and bond
percolation on the square, triangular and honeycomb lattices \cite{Sykes1964}.
A rigorous proof that $p_\mathrm{c} = 1/2$ for bond percolation on the square
lattice was given in \cite{Kesten1980}. A mapping was found
between a class of percolation models and corresponding models
of statistical physics, most notably the Potts model \cite{Fortuin1972}.
The values of several critical exponents as well as of
crossing probabilities \cite{Cardy1992} were rigorously established for
the site percolation on the triangular lattice \cite{Lawler2001,Smirnov2001ci,Flores2017},
and are believed to be universal for a wide class of planar
percolation models. Similar universality is believed to hold also
in higher dimensions, an important ingredient of advanced numerical methods
\cite{Newman2000,Newman2001,Feng2008,Wang2013,Xu2014, Koza2016}. Many nontrivial
properties of critical percolation clusters were also
derived using conformal field theory \cite{Pinson1994,Janssen2005,Simmons2007,Simmons2009}.

In numerical simulations of planar percolation rectangular systems
are preferred.
The crossing probability for such geometry is
defined for $p=p_\mathrm{c}$ as the probability that there exists
a percolating cluster that spans two opposite sides of the rectangle.
Pruessner and Moloney~\cite{Pruessner03} considered also the probability
that a cluster spans three sides of a rectangle: two long and a short one.
Using extensive simulations,
they conjectured that this probability, which we denote as $p_3$,
tends to 1/2 as the rectangle's aspect ratio $r$ diverges to infinity,
but gave no justification for this limit.
This conjecture raises several questions. Could such a simple result be exact?
If so, to what extent it is universal?
In particular, is it valid only in  the thermodynamic limit or perhaps
it is also valid for some finite lattices?

To answer these questions,
we start from numerical analysis of the system considered in \cite{Pruessner03};
however, to get the thermodynamic limit
we use extrapolation of finite-size results rather than
rely on simulations of large systems.
We study site percolation on a square lattice restricted to an elongated rectangle
of height $H$ and length $L$ (lattice units), with $L \gg H$.
Using the 64-bit Mersenne-Twister random number generator \cite{Matsumoto1998}
we generate a sequence of $L$ columns of height $H$, keeping in the computer memory
only the last two of them. These columns have their sites marked as occupied with probability
$0.592~746~050~792~10$, the best known value of $p_\mathrm{c}$ for the site percolation
on the square lattice \cite{Jacobsen2015}.
The occupied sites form clusters---we identify them with the union-find algorithm
\cite{Hoshen76,Newman2001}, assuming free boundary conditions on all system's edges.
While adding columns, we update some bits of information (specified below)
related to the clusters.
Even though eventually we are interested only in percolating
clusters (i.e., those spanning the two longer sides),
all clusters are monitored, because it is not known beforehand whether a given cluster
will develop into a percolating one. This contrasts to the method used
in~\cite{Deng2005}, where a special cluster labeling technique was developed
and the information on only the clusters reaching the last added column was kept in memory.
As soon as $L$ columns have been added, we store (append) in a disk file
the information about those of the clusters that percolate.
This is repeated until a preset number of percolating clusters has been generated for given $H$.

\begin{figure}
\includegraphics[width=0.7\columnwidth]{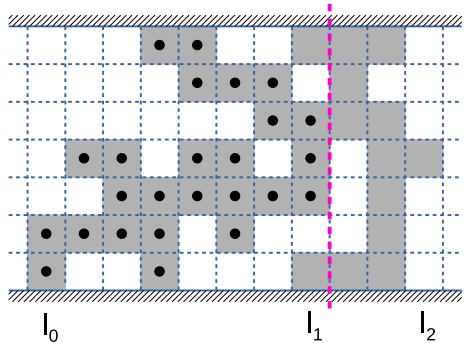}
\caption{\label{fig:def-params}An exemplary percolation cluster for $H=7$.
Parameters $l_0$ and $l_2$ are defined as the column numbers of its leftmost
and rightmost sites, respectively, and $l_1$ is the smallest column number
such that the cluster between $l_0$ and $l_1$ percolates.
The dots mark this minimal percolating cluster.
The dashed line marks a line splitting the system into two.}
\end{figure}

For each percolating cluster $i$ we store four integer parameters:
its mass (i.e., the number of the sites it occupies) $m(i)$ and parameters $l_0(i)$, $l_1(i)$,
and $l_2(i)$, as defined in the caption for Fig.~\ref{fig:def-params}.
We then define the cluster width $w(i)\equiv l_2(i) - l_0(i) + 1$,
and the gap between two consecutive clusters, $g(i) \equiv l_0(i+1) - l_2(i) -1$.
Clearly, $w(i) \ge 1$, whereas $g(i)$ can be positive, zero or negative.
Their average, in the limit of $L\to\infty$, will be denoted as
$w_H$ and $g_H$, respectively.

We also define two  additional integer parameters: $\gamma^+(i) \equiv l_2(i) - l_1(i) + 1$
and $\gamma^-(i) \equiv l_1(i) - l_2(i-1) -1$.
If one cuts vertically the system along any of the columns contributing to $\gamma^+(i)$,
so that this column becomes the right-hand-side edge of the system,
then cluster $i$ will be touching at least three sides of the rectangle:
top, bottom and right~(Fig.~\ref{fig:def-params}).
Since only one cluster can touch three given sides of a rectangle,
the intervals defining $\gamma^+(i)$ are mutually disjoint.
Hence, if we neglect the vicinity of the system's vertical borders or consider
an infinite system along the horizontal direction,
any column contributes exactly once to either $\gamma^+$ or $\gamma^-$
for some cluster~$i$.
Let $\gamma^\pm_H$ be the average of $\gamma^\pm(i)$ over
all percolating clusters $i$ in the limit of $L\to\infty$. Then
\begin{equation}\label{eq:p3(H)}
p_3(H) \equiv \frac{\gamma^+_H}{\gamma^+_H + \gamma^-_H}
\end{equation}
is the probability that if in a horizontally infinite strip of hight $H$ one randomly selects
a column and splits the system into two semi-infinite ones so that this column becomes
the right-hand-side edge of a semi-infinite rectangle,
this column will contain at least one site belonging to a vertically
percolating cluster.
For example, if one cuts the system shown in Fig.~\ref{fig:def-params}
along the dashed vertical line, the cluster that was percolating
in the original infinite system would still be percolating and this line
would contribute to $\gamma_H^+$ and, consequently, to $p_3$.
Selecting this line anywhere between $l_1(i)$ and $l_2(i)$ would have the same effect.
However, if the cutting line was selected to the left of $l_1$(i), down  to $l_2(i-1)$,
it would split the cluster into at least two non-percolating ones or even
miss any percolating cluster, and such cuttings would contribute
to $\gamma_H^-$, or $1 - p_3(H)$.

We performed the simulations for systems with  $2 \le H \le 1000$ and $L/H$
varying from $\approx 400$ for $H = 1000$ to $\approx 10^7$ for $H=10$.
The value of $H$ is limited by the computation time, which is of order of $H^2$
per percolating cluster, whereas $L$ is restricted by the available computer
memory per CPU core.
The number of clusters generated in our simulations varied between
$\approx 2\times 10^9$  for $H \ge 100$ and $\gtrsim 5 \times 10^{10}$ for $H< 100$.

\begin{figure}
	\includegraphics[width=0.95\columnwidth]{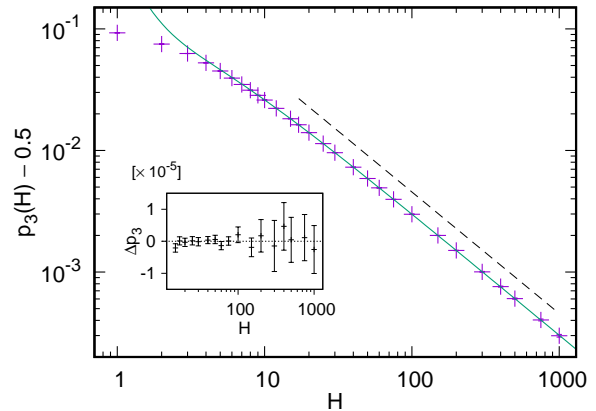}
	\caption{\label{fig:p3} Convergence of $p_3$ to $1/2$ as a function of the system height, $H$.
	The solid line is a fit to (\protect\ref{eq:p3-scaling}) for $H\ge 17$, and the dashed line is a guide
		to the eye with the slope $-1$.
		Inset: the difference between the simulation data and the fit.}
\end{figure}

In Fig.~\ref{fig:p3}
we present a log-log plot of $p_3(H) - 1/2$ as a function of $H$.
It suggests that $p_3(H)$ actually converges to $1/2$, as conjectured
in~\cite{Pruessner03}, and the convergence is of power-law type,
$p_3(h)-1/2 \propto  L^{-\omega}$ with $\omega \approx 1$.
To investigate the convergence, we approximate $p_3$ using
\begin{equation}
\label{eq:p3-scaling}
p_3(H) - 1/2 \approx  a_0 + a_1 H^{-1} +\ldots+ a_{M}H^{-M}.
\end{equation}
Some authors use a similar formula with an additional term $\propto L^{-\omega}$,
where $\omega$ is a non-integer correction-to-scaling exponent
\cite{Ziff2011,Xu2014,Wang2013}.
However, we found no evidence that such correction is necessary for $p_3$.
Fitting our data to (\ref{eq:p3-scaling}) with $M=3$ we found that the minimum value of
the regression standard error, $s = \sqrt{\chi^2/\mathrm{dof}}$
(the square root of the chi-squared statistic per degree of freedom),
is obtained if the fit is performed for $H \ge H_\mathrm{\mathrm{min}} = 17$,
in which case $s \approx 0.6$.
We used $ H_\mathrm{min}$ as the lower bound for the fits.
In this way we obtained
$a_0 = 1.4(15) \times 10^{-6}$, $a_1=0.3035(2)$, $a_2 = -0.499(6)$, and  $a_3=0.67(6)$.
This is a strong evidence that $p_3(H)$ indeed tends to $1/2$ as $H\to \infty$,
with the uncertainty of this limit, $1.5 \times 10^{-6}$, being
two orders of magnitude smaller than $4\times 10^{-4}$ reported in \cite{Pruessner03}.

The values of $p_3$ can be readily determined analytically for $H=1$ and $2$:
$p_3(1) = p_\mathrm{c}$ and
$p_3(2) = p_\mathrm{c}^2(1 + p_\mathrm{c} - p_\mathrm{c}^2)/(
                                  1 - p_\mathrm{c} + p_\mathrm{c}^2)$.
Our numerical result for $p_3(2)$, $0.574~954~8(12)$, is consistent with
this formula, which indirectly validates our numerical procedure.

Incipient percolating clusters are known to be fractals
with their mass scaling as $\lambda^{d_\mathbf{f}}$, where $\lambda$
is a characteristic length and $d_\mathbf{f}$ is the fractal dimension,
$d_\mathbf{f} = 91/48$.
For the model considered here, $\lambda \equiv H$ and so we expect
$\tilde{m}_{H} \equiv m_H/H^2 \propto H^{-5/48}$.
This can be used to do an additional check of correctness of our simulations.
We fitted our data to
\begin{equation} \label{eq:fit_with_b}
\tilde{m}_{H} \approx H^{-b}(a_0 + a_1/H + a_2/H^2),
\end{equation}
obtaining $b=0.1050(4)$, which agrees with the expected value
$5/48\approx 0.1042$.
The uncertainty of $\tilde{m}_{H}$ is significantly higher
than that found for $p_3$.
The fit is rather poor, with the regression standard error $s$ being
of order of 10 (included into the uncertainty estimation).
However, previous attempts to determine $d_\mathrm{f}$
numerically showed that the convergence of $\tilde{m}_{H}$ to the limit
$H\to\infty$ is slow and the uncertainties reported there were
even larger than ours~\cite{Kapitulnik1983,Hoshen1997}.

\begin{figure}
	\includegraphics[width=0.85\columnwidth]{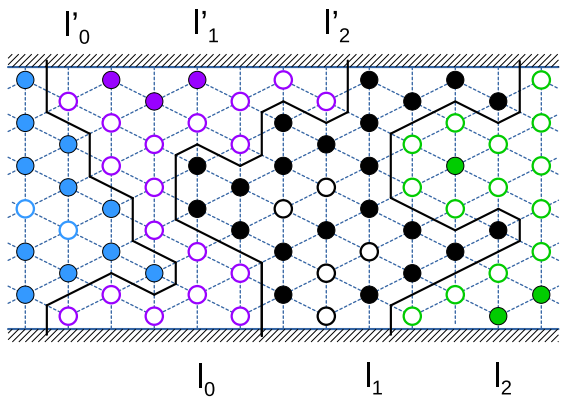}
	\caption{\label{fig:triangular}
		A distribution of occupied (filled circles)  and unoccupied (empty circles) sites on a triangular
		lattice with $H=6$. Solid lines mark the boundaries between alternating regions dominated
		by occupied and unoccupied sites. Parameters $l_0$, $l_1$, and $l_2$ characterize
		the percolating cluster marked with black filled circles, whereas  $l'_0$, $l'_1$, and $l'_2$
		refer to the left-adjacent cluster of unoccupied sites.}
\end{figure}

Just as we consider clusters of occupied sites, we can also investigate
clusters of unoccupied sites. A finite cluster of one species
(occupied or unoccupied) is surrounded by sites occupied by the other species.
Triangular lattice has a special property that all these surrounding sites
are connected and hence belong to the same cluster.
This property suffices to show that at any configuration, either occupied
or unoccupied sites percolate and to predict
the value of the percolation threshold, $p_\mathrm{c} = 1/2$ \cite{Sykes1964}.
Moreover, if we draw a line separating a cluster of occupied sites from unoccupied ones,
all sites on one of its sides will belong to this cluster,
whereas all sites on the other side will belong to another, single
cluster of empty sites (Fig.~\ref{fig:triangular}).
Thus, a rectangular system at or near $p_\mathrm{c}$ is composed of alternating
regions dominated by occupied or unoccupied sites.
The border of each such region is determined by the border
of a percolating cluster, which may surround
nonpercolating clusters of the opposite species, which, in turn, may contain
nonpercolating clusters of the original species and so on.
A region dominated by a percolating cluster of occupied or unoccupied sites
will be called ``conductive'' or ``nonconducting'', respectively.

Let $l_0$, $l_1$, and $l_2$ characterize a cluster of occupied sites and
$l'_0$, $l'_1$, and $l'_2$ --- the preceding cluster of empty sites
(filled black and open magenta circles in Fig.~\ref{fig:triangular}, respectively).
Recall that $l_1$ is the $x$ coordinate
of the first column such that if the cluster is cut vertically right after
column $l_1$, the cluster percolates and touches three edges of the system.
For the triangular lattice, $l_1$ is the maximum value of the $x$ coordinate
of the sites adjacent to its left-hand side border. This, in turn implies
\begin{equation}
\label{eq:l1=l'2+1}
l_1 = l'_2 + 1,
\end{equation}
because $l'_2$ is the largest $x$ coordinate of the
empty sites adjacent to the same border from the other side.
Thus, to determine the values of $l'_2$ and $l_1$, it suffices to
consider only the sites adjacent to the line separating the two clusters.
Plugging (\ref{eq:l1=l'2+1}) into the denominator of (\ref{eq:p3(H)}) yields
$\langle l_2(i) - l_1(i) + 1 \rangle + \langle l'_2(i) - l'_1(i-1) +1 \rangle$,
where $\langle \ldots \rangle$ denotes the average over the clusters ($i$)
in the thermodynamic limit.
However, the occupied and unoccupied regions must have identical
statistical properties at $p_\mathrm{c} = 1/2$, hence the two averages
must be equal to each other.
This implies that the site percolation on the triangular lattice is
characterized by
\begin{equation}\label{eq:p3(triangular)}
p_3(H) = 1/2
\end{equation}
for any $H$.
This can be independently verified for $H = 1$, as then
$p_3 = p_\mathrm{c}$, and for $H=2$, in which case,
after some cluster counting,
$p_3 = (2-p_\mathrm{c})p_\mathrm{c}^2/(1-p_\mathrm{c} + p_\mathrm{c}^2) = 1/2$.

\begin{figure}
	\includegraphics[width=0.85\columnwidth]{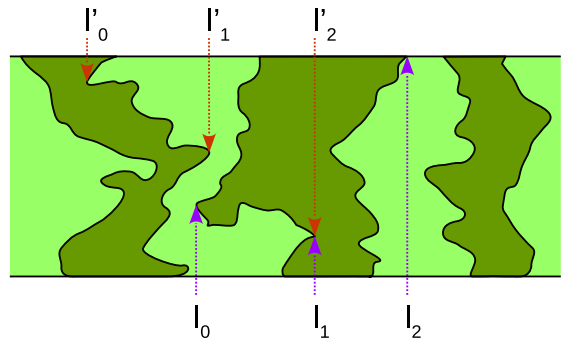}
	\caption{\label{fig:any}
		Percolation on an infinite (planar) strip generates alternating regions
		``dominated'' by occupied (dark green) and
		unoccupied (light green) regions. We conjecture that at $p_\mathrm{c}$
		for any periodic lattice
		$\langle l_2-l_1\rangle/\langle l'_2-l'_1\rangle \to 1$
		as $H \to \infty$.
	}
\end{figure}

While equation (\ref{eq:p3(triangular)}) was derived for the triangular lattice,
it sheds a new light onto the case of lattices that are not self-matching,
including the square lattice.
An infinite strip on a periodic lattice at (or near)
$p_\mathrm{c}$ can be divided into altering conductive and nonconducting
regions~(Fig.~\ref{fig:any}).
Let us for a moment consider the square lattice.
Just as for the triangular lattice, one can define the line separating
two consecutive regions. The occupied sites on the border of the conductive phase
will belong to the same percolating cluster. The situation in the
nonconducting regions appears to be more complex unless one realizes that
the unoccupied sites on the square lattice may be considered as ``connected''
directly if they are nearest (NN) or next-neighbor neighbors (NNN).
Then, the conductive regions will be controlled by standard site percolation
of occupied sites on the square lattice,
whereas the nonconducting ones---by site percolation of unoccupied sites
on the square lattice with NN and NNN neighbors.
Within this interpretation, (\ref{eq:l1=l'2+1}) will still hold,
but (\ref{eq:p3(triangular)}) can no longer be taken for granted,
as there is no trivial symmetry between the square NN and NN + NNN lattices.
Our numerical results (Fig.~\ref{fig:p3}) suggest that for the square
lattice Eq.~(\ref{eq:p3(triangular)}) is valid only in the limit of $H\to\infty$:
\begin{equation}\label{eq:H-to-infty-is-0.5}
p_3(H) \to 1/2 \quad \mathrm{as} \quad H\to \infty.
\end{equation}

Simplicity of this limit suggests some deeper relation between the
conducting and unconducting phases.
Indeed, the square lattices with NN  and NN + NNN connections form
a pair of mutually matching graphs \cite{Sykes1964} and we conjecture that
this can ba generalized to other periodic lattices: the nonconducting phase can be
understood as being built on the lattice matching the lattice of the
conductive phase.
This is in agreement with the observation that the percolation thresholds
of a lattice and its matching lattice always sum up to 1 \cite{Sykes1964}.

As the system size goes to infinity, the line separating the two phases
gets more and more complicated, eventually forming a fractal \cite{Saleur1987}.
If this line  is conformally invariant in the continuous limit
then it can be described as a stochastic Loewner evolution SLE$_6$~\cite{Schramm2000,Bauer2006}.
Validity of this assumption was shown rigorously for the triangular
lattice \cite{Smirnov2001ci}.
Since the hull of a critical percolation cluster is believed to be
conformally invariant irrespective of the underlying lattice, we expect SLE$_6$
to be a universal description of the lines separating occupied and unoccupied
``phases'' in the limit of $H\to\infty$. This limiting line is symmetric
in the sense that one cannot tell which of its sides is taken by a cluster
of ``occupied'' sites, and which by ``unoccupied''. Therefore we expect that
the alternating regions of occupied and unoccupied ``phases'' are statistically
indistinguishable in the limit of $H\to\infty$. This, in turn, leads to
(\ref{eq:H-to-infty-is-0.5}) as a universal property of percolation on planar lattices.
For systems built on self-matching lattices this line is symmetric
also for finite system sizes,
which implies the stronger condition (\ref{eq:p3(triangular)}) to be valid
for all self-matching lattices.
This symmetry, however, is broken for finite spanning clusters built
on a lattice that is not self-matching. As can be easily verified e.g.\
for the square lattice, the lines separating different, non-matching
phases are asymmetric with respect to the possibility of self-touching
(or forming loops without self-cutting) and touching of two adjacent lines.
This explains why $p_3(H)\neq1/2$ for such systems.

The nature of $p_3$ is similar to the crossing and wrapping probabilities
\cite{Cardy1992,Pinson1994},
except that it is a three-leg quantity, i.e.,
it describes an event of a cluster touching three system sides. Moreover, it
involves conductive and nonconducting ``phases'' (Fig.~\ref{fig:any})
rather than individual clusters. While in reality the borders of these phases
are complex fractals,
we neither needed this information nor can infer it from our arguments and actually can treat the phases as
thick, almost classical objects that are potentially less sensitive to finitie-size effects than
rarefied critical percolating clusters.
We also found no evidence that $p_3$ is related to any nonclassical (non-integer)
exponent; still, just as the crossing and wrapping probabilities
\cite{Newman2000,Newman2001,Feng2008,Wang2013,Xu2014},
whose values have been derived using advanced field-theoretical methods, it could be used
as a sensitive tool  for localization of $p_\mathrm{c}$ when $p$ is
varied around it.

In summary, we have presented evidence that the probability $p_3$ that there exists
a cluster touching the three sides of a semi-infinite strip
at the percolation threshold $p_\mathrm{c}$ has a universal continuous limit: 1/2.
The convergence of $p_3$ for the site percolation on the square lattice at  $p_\mathrm{c}$
is quick and robust, which suggests that it may be an effective
parameter for determining $p_\mathrm{c}$ and related parameters numerically also for other planar lattices.
We attribute the value of the limit, 1/2, to the symmetry of the separation lines
between ``conductive'' and ``nonconducting'' phases, which are believed to be
universally described by the SLE$_6$ process.

We acknowledge helpful discussions with Grzegorz Kondrat.


%

\end{document}